\documentclass{ws-procs9x6}            
\def\note #1]{{\bf #1]}}
\def\fig{.}
\newcommand{\etal}{{\it et~al.}}
\newcommand{\cf}{{\it cf.}}
\newcommand{\eg}{{\it e.g.}}
\newcommand{\aap}{{\it Astron.\ Astrophys.}}
\newcommand{\apj}{{\it Astrophys.\ J.}}
\newcommand{\apjl}{{\it Astrophys.\ J.}}
\newcommand{\apjs}{{\it Astrophys.\ J.\ Suppl.}}
\newcommand{\aplett}{{\it Astrophys.\ Lett.}}
\newcommand{\apss}{{\it Astrophys.\ Space Sci.}}
\newcommand{\araa}{{\it Annu.\ Rev.\ Astron.\ Astrophys.}}

\newcommand{\mnras}{\it Mon.\ Not.\ R.\ ast.\ Soc.}
\newcommand{\physrep}{\it Physics Rep.}
\newcommand{\ssr}{\it Space Science Rev.}
\newcommand{\solphys}{\it Solar Phys.}
\newcommand{\nat}{\it Nature}
\newcommand{\CG}{{\cal G}}
\newcommand{\CI}{{\cal I}}
\newcommand{\CK}{{\cal K}}
\newcommand{\dd}{{\rm d}}
\begin{document}
\title{Helioseismology and solar neutrinos}

\author{J{\o}rgen Christensen-Dalsgaard$^*$}

\address{Stellar Astrophysics Centre, Department of Physics and Astronomy,\\
Aarhus University,\\
8000 Aarhus C, Denmark\\
$^*$E-mail: jcd@phys.au.dk}

\begin{abstract}
The studies of solar neutrinos and helioseismology have been closely intertwined
since the first neutrino experiment and the first observations of solar
oscillations in the sixties.
Early detailed helioseismic analyses provided strong support for the standard
solar model and hence a clear indication that the solution to the
discrepancy between the predicted and observed neutrino fluxes had
to be found in terms of neutrino physics, as now fully confirmed by
direct observations.
With the full characterization of neutrino properties we are now in a position
to combine neutrino observations and helioseismology to obtain a more
complete understanding of conditions in the solar core.
Here I provide a personal and largely historical overview of these developments.
\end{abstract}


\bodymatter

\section{Introduction}
\label{sec:intro}

The possibility to detect neutrinos from the nuclear reactions in the solar
core clearly provided a very exciting potential for testing otherwise
inaccessible parts of a star, including providing the definite confirmation
that stars like the Sun derive their energy from hydrogen fusion.
Thus it was a major problem that the early upper limit on the neutrino
flux determined by Davis {\etal}\cite{davisetal68} was lower by around
a factor of seven than the predictions of then up-to-date solar models, 
such as the model by Bahcall {\etal}\cite{bahcalletal68}.
This clearly raised doubts about the understanding of stellar structure and
evolution, with potentially serious consequences for many areas of 
astrophysics.
There was an obvious need for other observations that might probe conditions
in the solar core.

Such observations have been provided through the study of oscillations
on the solar surface, in what is now known as {\it helioseismology}.
Since the middle of the seventies this field has developed in parallel with
the solar neutrino investigations.
One can reasonably claim that by the early nineties the helioseismic
constraints on the structure of the solar core essentially had eliminated
the models proposed to reduce the neutrino flux to the observed level,
strongly pointing towards a solution of the solar neutrino problem
in terms of non-standard neutrino physics, as was later confirmed
by the direct observation of neutrino flavour oscillations.
Given the resulting understanding of the neutrino properties,
and new observations of solar neutrinos, the measured neutrino fluxes can
now be used to probe the solar interior, as an important complement to
helioseismology.
As was clear from this conference, the prospects for these investigations
are excellent, given the upcoming ambitious new experiments.

Here I provide a largely historical, and personally biased, overview of the
relation between neutrino studies and helioseismology, which has been a central
issue throughout my career, since my early days as a PhD student in Cambridge.

General reviews on helioseismology were provided in
Refs~\citenum{cd02} and \citenum{aertsetal10},
while Ref.~\citenum{cd04} provided a more detailed presentation of the
history of helio- and asteroseismology.
An extensive discussion of the early phases of the neutrino studies
was provided by Bahcall\cite{bahcall89}, while Haxton\cite{haxtonetal13}
presented the situation of only a few years ago.
The present status is outlined in a number of papers in the present proceedings,
including the review by Francesco Villante.

\section{The solar spoon and early helioseismology}
\label{sec:heliostart}

The discrepancy between the predicted and observed neutrino flux was a strong
indication of problems with solar modelling.
Since the Davis experiment was sensitive predominantly to the high-energy
neutrinos resulting from the decay of ${}^8{\rm B}$, with a dependence
on the central temperature of the Sun to a high power,
the modifications of the solar models involved a reduction of the central
temperature, while maintaining the observed total energy flux from the
Sun.
It was pointed out by, for example, Ezer \& Cameron\cite{ezer_cameron68}
that this could be achieved by mixing the core of the Sun, hence increasing
the central hydrogen abundance and thus reducing the core temperature 
required to generate the solar luminosity.
Dilke \& Gough\cite{dilke_gough72} proposed a creative variant of this model%
\footnote{in a paper entitled `The solar spoon'}:
this involved intermittent instability of the solar core to standing internal
gravity waves (or g modes) which through nonlinear development caused 
onset of core convection and hence the mixing.
Dilke \& Gough postulated that the Sun had undergone such an episode within
the last few million years, largely switching off the nuclear reactions,
and that it was now recovering on a thermal timescale.
A result of this process was a decrease in the solar luminosity by a few
per cent which, they suggested, was related to the ongoing series of
glaciations.

This proposal also set the scene for the start of my PhD studies in Cambridge
in the fall of 1973, with Douglas Gough as my supervisor.
As a first project I completed the detailed analysis of the stability of
solar models started by Fisher Dilke as part of his PhD studies.
The results\cite{cdetal74} did indeed confirm the assumed instability;
this arose from the build-up of a steep gradient in the abundance
of ${}^3{\rm He}$ in the core, initially over a period of about 200 Myr.
Following the idea of Dilke \& Gough the resulting mixing would therefore
suppress the instability until a renewed gradient had been established,
and hence leading to the postulated series of intermittent mixing episodes.

It was never demonstrated that the instability, which has a growth time of
order $10^7$ years, would in fact lead to convective instability and
hence mixing.
Even so, starting from this analysis I continued work on improved solar
modelling and investigations of stability analysis.
This provided a basis for the involvement in helioseismology, as the field
got underway.

Oscillations in the solar atmosphere with periods near five minutes
were identified by Leighton in 1961
(see Ref.~\citenum{leightonetal62}) but were generally considered to
be atmospheric phenomena.
However, Ulrich\cite{ulrich70} and Leibacher \& Stein\cite{leibacher_stein71}
proposed that they might in fact be standing acoustic waves, 
trapped in the outer parts of the solar interior.
This was confirmed by the observations of Deubner\cite{deubner75},
with sufficient temporal and spatial resolution to
show a clear modal structure of the oscillations, as a function
of frequency and wavenumber.
These results provided a way to use the observed properties of the
oscillations to investigate the outer parts 
of the Sun \cite{gough77, ulrich_rhodes77, deubneretal79}.

While the modes observed by Deubner did not provide direct information
about the deep solar interior, 
such information was promised by solar oscillations with a period of
160 min, announced in 1975 by Brookes {\etal}\cite{brookesetal76}
and Severny {\etal}\cite{severnyetal76}:
such oscillations could only be due to g modes of fairly high radial order,
which are very sensitive to conditions in the solar core.
Perhaps even more exciting was the announcement by Henry Hill in 
June 1975 at the IBM conference on Astrophysical Fluid Dynamics,
organized by Douglas Gough in Cambridge, 
of evidence of several oscillations in the solar diameter
(see Refs \citenum{hilletal75, hilletal76, hill_caudell79});
these were the serendipitous result of very careful observations
of the solar oblateness\cite{hilletal74}.
With several modes of oscillation that appeared to be of a global nature
this promised relatively detailed information on the solar interior.
Given my ongoing work on solar models and oscillations I computed
a set of frequencies for a solar model and presented a comparison with
the observations the following day, showing apparently good agreement
between the observed frequencies and the model.

Later work has shown that the 160 min oscillation
and the signal observed by Hill and his group had no connection
to global solar modes but were likely caused by fluctuations in the Earth's
atmosphere.
Even so, these early results for many served as a kick-off for 
studies of the promise of global solar oscillations as a diagnostics of
the solar interior\cite{scuflaireetal75, cd_gough76, iben_mahaffy76}.

\section{Large-scale five-minute oscillations}

Truly global solar modes of oscillation were announced in 1979 by the
group in Birmingham led by George Isaak,
based on observations carried out on Tenerife\cite{claverieetal79}.
As did the modes observed by Deubner they had periods near five minutes,
but since the observations were carried out in light integrated over
the solar disk, smaller-scale oscillations were filtered out by averaging.
Global solar modes have the structure of spherical harmonics
$Y_l^m(\theta, \phi)$ as functions of co-latitude $\theta$ and
longitude $\phi$, 
and the observations were interpreted\cite{cd_gough80b, cd_gough82}
as resulting from high-order acoustic modes of 
spherical-harmonic degree $l = 0 - 3$.
The cyclic frequencies $\nu_{nl}$ of such modes, $n$ being the radial order,
satisfy the asymptotic relation\cite{tassoul80}
\begin{equation}
\nu_{n\,l} \simeq \Delta\nu \left( n+\frac{l}{2}+\epsilon \right)
-d_{n\,l} \; ,
\label{eq:asymp}
\end{equation}
where
\begin{equation}
\Delta\nu = \left(2 \int_0^R \frac{{\rm d}r}{c} \right )^{-1} \; ,
\label{eq:dnu}
\end{equation}
$r$ being the distance to the centre, $R$ the surface radius of the star
and $c$ the adiabatic sound speed;
here $\epsilon$ is a phase largely determined by conditions near the stellar
surface and $d_{nl}$ is a smaller second-order term.
Thus the basic structure of the observed spectrum consists of modes
of a given degree separated by the {\it large frequency separation}
$\Delta \nu$,
with modes of odd degree located halfway between the adjacent modes
of even degree.
To leading order the frequencies $\nu_{nl}$ and $\nu_{n-1\,l+2}$ coincide;
this was indeed the structure observed in Ref.~\citenum{claverieetal79}.
Including the second-order term leads to the {\it small frequency separation}
\begin{equation}
\delta \nu_{nl} = \nu_{nl} - \nu_{n-1\,l+2}
\simeq -(4l+6)\frac{\Delta\nu}{4\pi^2 \nu_{n\,l}}
\int_0^R\frac{{\rm d} c}{{\rm d} r}\frac{{\rm d} r}{r}. \; 
\label{eq:smallsep}
\end{equation}
Owing to the scaling by $r^{-1}$ this is heavily weighted towards 
the solar centre and hence is sensitive to conditions in the solar core;
the potential importance for elucidating the origin of the
solar neutrino problem is obvious (see also \fref{fig:neutweight}).

\begin{figure}
\begin{center}
\includegraphics[width=4.5in]{\fig/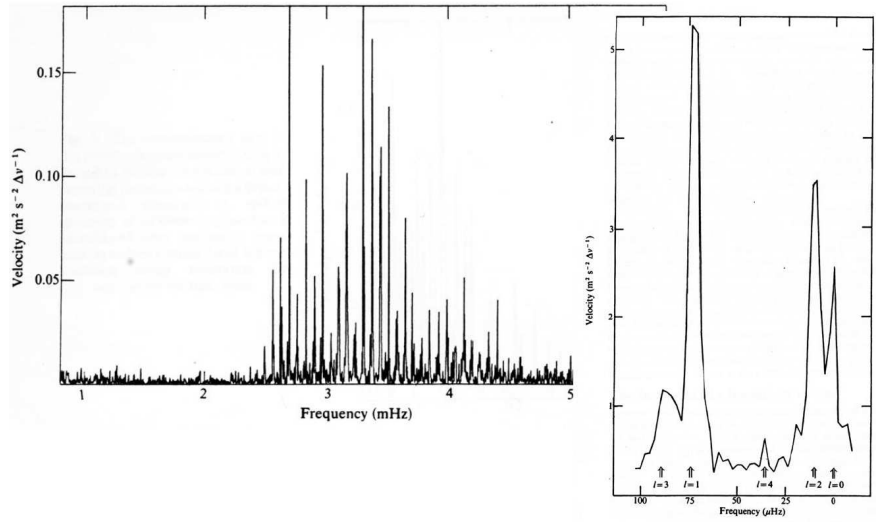}
\end{center}
\caption{Power spectrum of solar oscillations in light integrated over the
solar disk, from continuous radial-velocity observations over nearly 
six days at the geographical South Pole.
The right-hand panel shows a collapsed spectrum, obtained by summing
segments of length $\Delta \nu$ ({\cf} Eq.~\ref{eq:asymp})
and showing the pairs of $l = 3, 1$ and $l = 2, 0$ modes.
Adapted from Grec {\etal}\cite{grecetal80}.
}
\label{fig:grec}
\end{figure}

The full asymptotic behaviour was dramatically demonstrated by the results,
illustrated in \fref{fig:grec},
of almost continuous observations over six days 
by Grec {\etal}\cite{grecetal80} from the geographical South Pole.
This clearly showed a broad range of modes of degree 0 to 3, with a well-defined
envelope of amplitude;
the averaged fine structure, shown in the inset, confirms the ratio 3/5
predicted by \eref{eq:smallsep}
between $\delta \nu_{n0}$ and $\delta \nu_{n1}$.
Early analyses of these results ({\eg}, Ref.~\citenum{isaak80})
indicated that the Sun had a low abundance of helium and heavy elements
compared with standard models,
in conflict with Big-Bang nucleosynthesis, but yielding low neutrino fluxes. 
However, a more careful modelling of the observations\cite{cd_gough80}
showed that they were essentially consistent with the standard models.

\begin{figure}
\begin{center}
\includegraphics[width=3.5in]{\fig/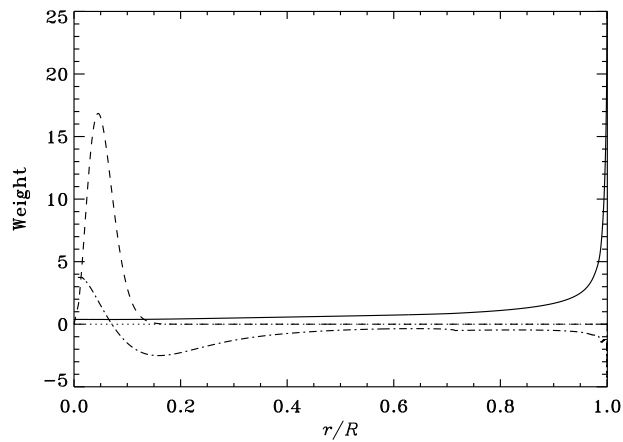}
\end{center}
\caption{Integrands with respect to $r$ for
the ${}^8{\rm B}$ neutrino flux, 
the large frequency separation ($c^{-1}$) and the
small frequency separation ($r^{-1} \dd c/\dd r$),
normalized such that the integrals (of the absolute value, in
the case of the small frequency separation) with respect to $r/R$ are one.
}
\label{fig:neutweight}
\end{figure}

Bahcall noted an important complementarity between the neutrino measurements
and helioseismology (see Fig.~4.3 of Ref. \citenum{bahcall89}): 
the neutrino fluxes depend on the structure of the core, while the
large frequency separation ({\cf} Eq.~\ref{eq:dnu}) is strongly
weighted towards the surface.
This is illustrated in \fref{fig:neutweight} showing,
both suitably normalized, $c^{-1}$ and 
the integrand $\CI_8$ defined such that the ${}^8{\rm B}$ neutrino flux
is proportional to $\int \CI_8 \dd r$.
However, Bahcall's argument neglected the diagnostics available in
the small separation ({\cf} Eq.~\ref{eq:smallsep});
as also illustrated in \fref{fig:neutweight} this has a substantial contribution
from the region dominating the neutrino flux,
providing a strong potential for investigating the origin of the neutrino
problem.

\begin{figure}
\begin{center}
\includegraphics[width=4.5in]{\fig/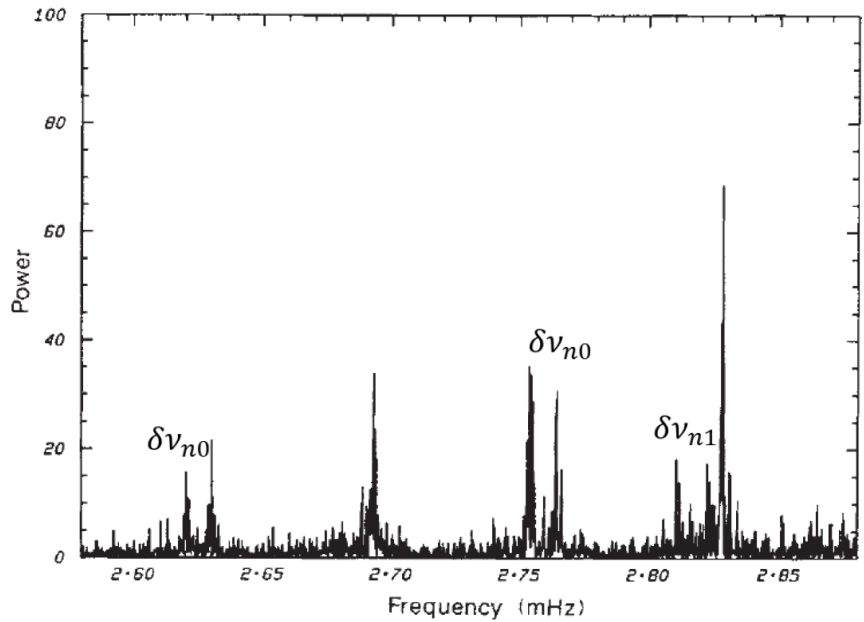}
\end{center}
\caption{Small part of power spectrum of solar oscillations, observed
in light integrated over the solar disk over two-month segments
from the Tenerife, Hawaii and Australia sites of the BiSON network.
Peaks defining the small separations $\delta \nu_{nl}$ 
({\cf} Eq. \ref{eq:smallsep}) are indicated.
Adapted from Elsworth {\etal}\cite{elsworthetal90}.
}
\label{fig:elsworth}
\end{figure}

These early promising results of helioseismology motivated strong efforts
to establish dedicated observing facilities to provide long and nearly
continuous observations of solar oscillations.
The Birmingham group expanded their observations to include sites also
in Hawaii and Australia, working towards the current six-station
BiSON%
\footnote{Birmingham Solar Oscillations Network}
network in operation since 1992\cite{chaplinetal96}.
In 1990 Elsworth {\etal}\cite{elsworthetal90} carried out a detailed
analysis of observations from these three stations.
\Fref{fig:elsworth} shows a small segment of the power spectrum of the
observations, clearly illustrating the expected frequency structure
and allowing accurate determination of the small separations.
In a seminal contribution to solar physics and the study of the 
neutrino problem Elsworth {\etal} compared the observations with
standard solar models, with a high neutrino flux, and two sets of 
models with reduced flux: one set included core mixing, as discussed above,
and the second included contributions to energy transport from the so-called
weakly interacting massive particles 
(WIMPs; see Refs \citenum{faulkner_gilliland85, spergel_press85}) which reduce
the temperature gradient in the interior of the model
and hence the central temperature,
leading to a reduced neutrino flux\cite{}.
The analysis was carried out in terms of linear fits to the small frequency
separation, of the form
\begin{equation}
\delta \nu_{nl} = d_l + s_l (n - n_0) \; ,
\label{eq:fitsep}
\end{equation}
with a reference order $n_0 = 21$.
The results, illustrated in a $(d_l, s_l)$ diagram,
strongly indicated that the standard models were essentially
consistent with the observations, whereas both the mixed and the WIMP
models were clearly inconsistent.
On this basis, Elsworth {\etal} concluded that

\begin{quote}
{\it ``Our results agree with standard solar models, and seem to remove the need
for significant mixing or weakly interacting massive particles (WIMPS)
in the core, both of which have been advanced to explain the low measured
flux of solar neutrinos. 
This suggests that the solar neutrino problem must be resolved within
neutrino physics, not solar physics;
neutrino oscillations and a finite neutrino mass form 
a possible explanation.''}
\end{quote}
Needless to say, this has now been fully confirmed.

\begin{figure}
\begin{center}
\includegraphics[width=4.5in]{\fig/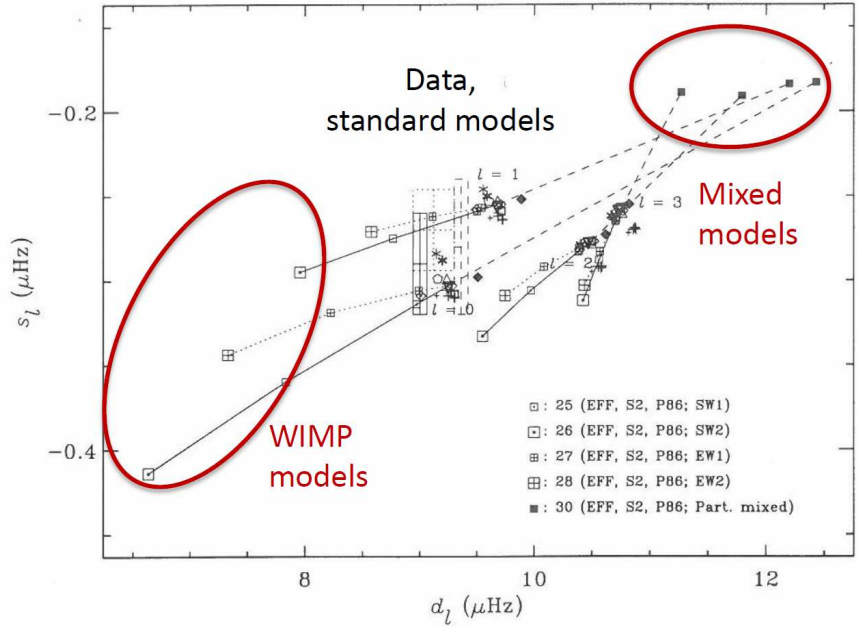}
\end{center}
\caption{Intercept $d_l$ and slope $s_l$ of the scaled small separations
fitted
as functions of mode order ({\cf} Eqs~\ref{eq:fitsep} and \ref{eq:scalesep}).
The solid and dashed rectangles show observations for $l = 0$ and
$1$ from Elsworth {\etal}\cite{elsworthetal90},
whereas the dotted rectangle shows an average between $l = 0$ and $1$
from Gelly {\etal}\cite{gellyetal88}.
Results are shown for several models and for $\hat \delta \nu_{nl}$
for $l = 0 - 3$ (see Ref.~\citenum{cd91} for details) including,
as indicated, models with reduced neutrino fluxes resulting
from WIMP-like modifications or core mixing.
Adapted from Ref.~\citenum{cd91}.
}
\label{fig:cdgafd91}
\end{figure}

A more extensive investigation of this nature\cite{cd91},
confirming the conclusions
of Elsworth {\etal}, is illustrated in \fref{fig:cdgafd91}; 
here the analysis was carried out in terms of the scaled small separation
\begin{equation}
\hat \delta \nu_{nl} = {3 \over 2 l + 3} \delta \nu_{nl} \; ,
\label{eq:scalesep}
\end{equation}
thus taking out the asymptotic $l$-dependence ({\cf} Eq. \ref{eq:smallsep}).

It should be noticed that it might have been possible to combine partial
mixing and the effect of WIMPs in such a way as to match the helioseismic
results while at the same time obtaining a reduced neutrino flux. 
However, such a model would obviously have been highly contrived.
A second important point in comparing the helioseismic and neutrino
results is that the acoustic-mode frequencies depend predominantly on the
sound speed, as is clear from the asymptotic expressions, 
Eqs (\ref{eq:dnu}) and (\ref{eq:smallsep}), while the neutrino flux
is predominantly sensitive to temperature.
Approximating matter in the solar core by a fully ionized ideal gas the
relation between sound speed $c$ and temperature $T$ is
\begin{equation}
c^2 \simeq {5 \over 3} {k_{\rm B} T \over \mu m_{\rm u}} \; ,
\label{eq:csq}
\end{equation}
where $k_{\rm B}$ is Boltzmann's constant, $m_{\rm u}$ is the atomic mass unit
and $\mu$ is the mean molecular weight which depends on composition.
Thus helioseismology essentially constrains $T/\mu$ but not directly
$T$ and $\mu$ separately.
Even so, it is clear that the analysis by Elsworth {\etal} was a strong
support of the standard solar models and a clear indication that the
apparent solar neutrino deficit resulted from the neutrino physics, 
a decade before this was firmly established by direct observations
from the Sudbury Neutrino Observatory\cite{ahmadetal01, ahmadetal02}.

\section{Inferences of solar internal structure}
\label{sec:inv}

Although the low-degree modes provide crucial information about the solar core,
the potential of helioseismology is far wider.
The oscillations are excited at all spatial scales in a frequency range
around 3000 $\mu{\rm Hz}$, corresponding to roughly five minutes,
from the global modes detected by Claverie {\etal} to the small-scale,
and hence high-degree, modes observed by Deubner.
This property is the result of the excitation mechanism of the modes:
they are intrinsically stable but excited by the acoustic noise generated
by the near-surface convection.\cite{houdek02}
This results in excitation that is largely independent of spatial scale
and centred near periods of five minutes.
Since the Sun can be observed with high spatial resolution the full
range of modes can be detected, 
providing a huge wealth of data on the properties of the solar interior.
A crucial step were the observations by Duvall \& Harvey\cite{duvall_harvey83},
which established a link between the low- and high-degree observations,
securing an unambiguous identification of the observed modes.
These data also allowed the first helioseismic inference of the
solar internal sound speed\cite{cdetal85}.

\begin{figure}
\begin{center}
\includegraphics[width=2.5in]{\fig/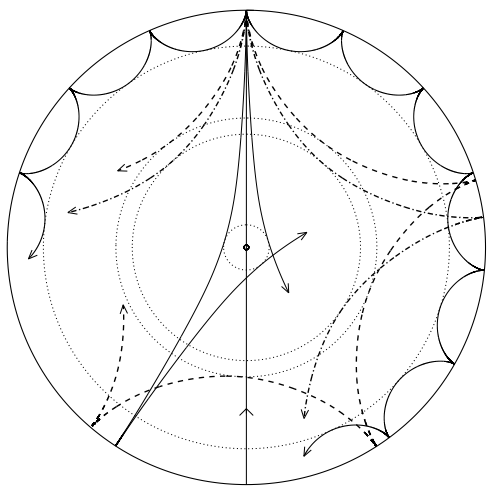}
\end{center}
\caption{Propagation of rays of acoustic waves in a cross-section of a model
of the Sun; the dotted circles indicate the location of the lower
turning points $r_{\rm t}$, defined by \eref{eq:rt}.
In order of increasing propagation depth the rays correspond to modes
of degree $l = 75, 25, 20$ and $2$, while the line going through the centre
schematically indicates the behaviour of a radial mode, with $l = 0$.
}
\label{fig:rays}
\end{figure}

The strength of the solar acoustic modes as probes of the solar interior 
is illustrated in \fref{fig:rays}, showing the properties of the modes
in terms of rays of sound waves.
When propagating from the surface towards the interior at an angle
the wave is refracted by the increasing sound speed, resulting from the
increase in temperature ({\cf} Eq. \ref{eq:csq}),
leading to a total internal reflection at a distance $r_{\rm t}$ from
the centre, determined by
\begin{equation}
{c(r_{\rm t})^2 \over r_{\rm t}^2} = {\omega^2 \over l(l+1)} \; ,
\label{eq:rt}
\end{equation}
where $\omega = 2 \pi \nu$ is the angular frequency.
As illustrated, low-degree modes therefore penetrate close to the centre
(and, in particular, provide information about the solar core,
as discussed above), while higher-degree modes are confined to a region
closer to the surface.
Modes covering the full range of degrees therefore effectively provide a
scan of the solar sound speed from the centre to the surface.
This behaviour was used explicitly in the asymptotic analysis in
Ref.~\citenum{cdetal85} but also underlies other techniques
for the so-called helioseismic
inverse analysis\cite{gough_thompson91, goughetal96}.

In its most precise form the analysis assumes that solar structure
is close to that of a reference model, 
such that the differences $\delta \omega_{nl}$ in frequency between
the Sun and the model can be linearized on the form
\begin{eqnarray}
{\delta \omega_{nl} \over \omega_{nl}} & \!\! = \!\! &
\int_0^R \left[ K_{c^2, \rho}^{nl}(r) {\delta_r c^2 \over c^2}(r)
+ K_{\rho,c^2}^{nl}(r) {\delta_r \rho \over \rho}(r) \right] \dd  r 
\nonumber \\
&& + Q_{nl}^{-1} \CG(\omega_{nl}) + \epsilon_{nl} \; ;
\label{eq:deltaomega}
\end{eqnarray}
here we used the adiabatic approximation for the computed frequencies,
such that they can be fully characterized by the sound speed and the
density $\rho$ in the model, and $\delta_r$ indicates differences
at fixed $r$.
The errors arising from the adiabatic approximation, and other inadequacies
in the modelling of the near-surface layers of the Sun,
are represented by the term in $\CG$, where $Q_{nl}$ is the mode inertia
normalized by the inertia of a radial mode of the same frequency,
and $\epsilon_{nl}$ is the observational error.
Also, the {\it kernels} $K_{c^2, \rho}^{nl}$ and $K_{\rho, c^2}^{nl}$
can be determined from the oscillation eigenfunctions computed from the
reference model\cite{gough_thompson91}.
The goal of the analysis is typically to obtain localized measures of the
difference in sound speed between the Sun and the model, as a function
of position $r_0$ in the Sun.
This is achieved by making linear combinations of Eqs~(\ref{eq:deltaomega})
with suitably chosen coefficients $c_{nl}(r_0)$, such that the
terms in $\delta_r \rho$ and $\CG$ are suppressed and the contribution
from the errors is constrained (for details, see Ref.~\citenum{rabelloetal99}).
If successful, we obtain a localized average of $\delta_r c^2$,
\begin{equation}
\overline{ \left({\delta_r c^2 \over c^2}\right)}(r_0) 
= \sum_{nl} c_{nl}(r_0) {\delta \omega_{nl} \over \omega_{nl}} \simeq 
\int_0^R  \CK_{c^2,\rho}(r_0, r) {\delta_r c^2 \over c^2} \dd r \; ,
\end{equation}
with a standard deviation which can be determined from the standard errors
of the observations, often assumed to be independent.
Here, the {\it averaging kernel}
\begin{equation}
\CK_{c^2, \rho}(r_0, r) = \sum_{nl} c_{nl}(r_0) K_{c^2, \rho}^{nl}(r) 
\label{eq:avker}
\end{equation}
provides a measure of the extent to which the estimate is localized near $r_0$.

\begin{figure}
\begin{center}
\includegraphics[width=4.5in]{\fig/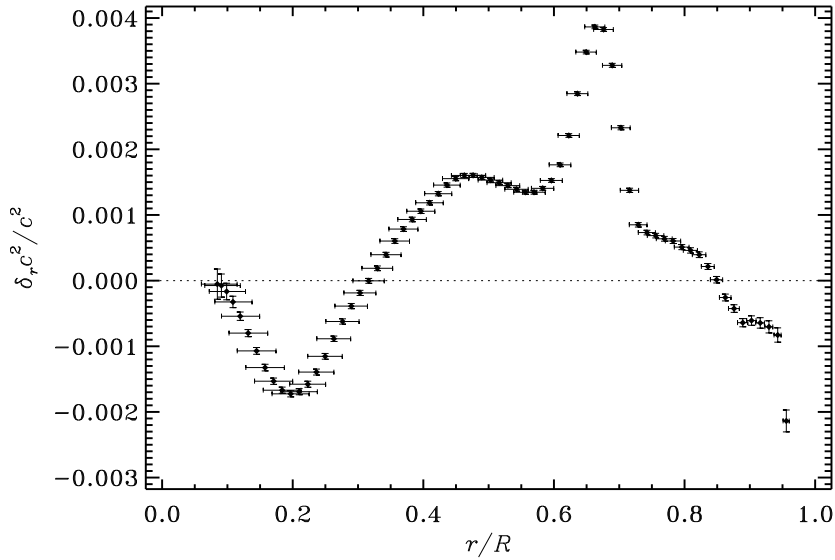}
\end{center}
\caption{Inferred differences in squared sound speed between 
the Sun and Model S\cite{cdetal96}, in the sense (Sun) -- (model),
inferred from inversion of observed solar frequencies.
The (barely visible) vertical bars indicate $1 \sigma$ errors in the inferences,
estimated from the errors in the observed frequencies,
whereas the horizontal bars provide a measure of the resolution 
of the inversion,
as determined by the averaging kernel ({\cf} Eq.~\ref{eq:avker}).
Adapted from Basu {\etal}\cite{basuetal96}.
}
\label{fig:csqinv}
\end{figure}

\begin{figure}
\begin{center}
\includegraphics[width=3.5in]{\fig/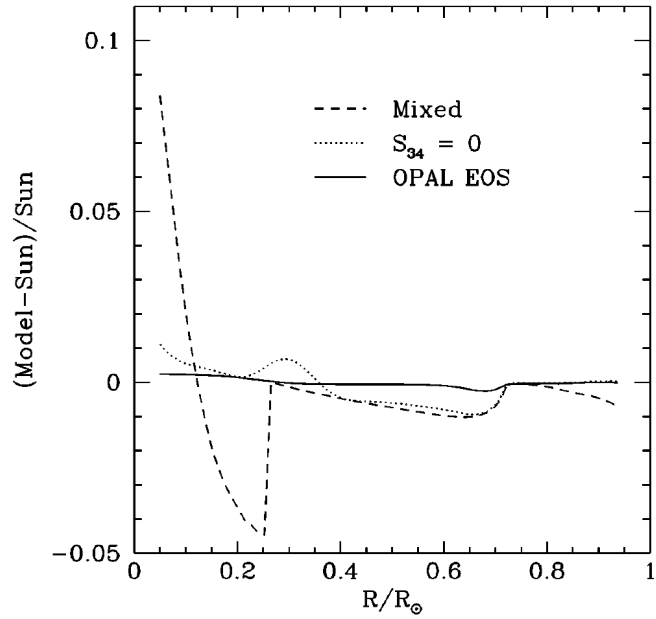}
\end{center}
\caption{Helioseismically inferred differences between three models and the Sun,
in the sense (model) -- (Sun).
The solid line shows a standard solar model, similar to the model illustrated
in \fref{fig:csqinv}, the dashed line a model with a mixed core and
the dotted line a model where the
${}^3{\rm He}+{}^4{\rm He}$ reaction has been switched off.
From Bahcall {\etal}\cite{bahcalletal97}.
}
\label{fig:bahcallinv}
\end{figure}

Differences in squared sound speed between the Sun and a typical model
from the mid-nineties are shown in \fref{fig:csqinv}.
The model, the so-called Model~S of Ref.~\citenum{cdetal96},
used essentially up-to-date physics of the time, including
diffusion and settling of helium and heavy elements.
Opacities were obtained from the OPAL tables\cite{rogers_iglesias92},
using the Grevesse \& Noels\cite{grevesse_noels93} heavy-element
composition with the ratio $Z_{\rm s}/X_{\rm s} = 0.0245$ between the
surface abundances by mass of heavy elements and hydrogen.
As shown by the horizontal bars the analysis is relatively successful
in providing localized measure of the difference in most of the Sun.
Also, although the model clearly by astrophysical standards provides
an excellent fit to solar structure,
the differences are far larger than the, barely visible, error bars.

The consequences of these results for the solar neutrino problem were
analysed by Bahcall {\etal}\cite{bahcalletal97},
as illustrated in \fref{fig:bahcallinv}.
In addition to a standard solar model similar to Model S they
considered two models with reduced neutrino fluxes:
a model with mixing in the core and a model where the
${}^3{\rm He}+{}^4{\rm He}$ reaction was switched off,
removing the high-energy neutrinos from the PP-II and PP-III branches
of the nuclear reactions.
The inferred differences for the standard model were, as for Model S,
very small.
The partially mixed model resulted in a huge difference,
to a large extent caused by the dependence of the sound speed on
the mean molecular weight ({\cf} Eq. \ref{eq:csq}),
but also the model with the modified nuclear reaction network showed
much larger departures from solar structure than did the standard model.
On this basis Bahcall {\etal} concluded that

\begin{quote}
{``\it [s]tandard solar models predict the measured properties
of the Sun more accurately than required for applications involving
solar neutrinos''},
\end{quote}
with the implied consequence that the neutrino flux
predicted by standard solar models is essentially correct;
and that therefore the solution to the neutrino problem had to be found in
neutrino physics.
These results clearly supported and strengthened the conclusion
reached by Elsworth {\etal}\cite{elsworthetal90}.

\section{A new solar problem?}

An important constraint on solar models is the ratio between the abundances
of elements heavier than helium and the abundance of hydrogen in the solar
atmosphere.
This can in principle be determined from spectroscopy.%
\footnote{Although helium can be observed in the solar spectrum from the
outer layers of the solar atmosphere, conditions are so complicated there 
that a reliable helium abundance cannot be inferred.}
As discussed by Nicolas Grevesse at the conference
(see also Ref.~\citenum{grevesseetal13}) substantial improvements have
taken place since around 2000 in the techniques used for 
abundance determinations.
These include the use of three-dimensional hydrodynamical simulations of
the solar atmosphere as basis for the analysis of the observed spectra,
rather than the simplified one-dimensional static models used in the past;
in addition, some account is taken of the departures from the assumption
of local thermodynamical equilibrium in the description of, for example,
the  distribution on ionization states and energy levels in the gas.
This has resulted in substantial reductions in the inferred abundances of, 
in particular, carbon, nitrogen and oxygen.
Early results were reviewed by Asplund\cite{asplund05}.
A convenient measure of the change in abundances is the ratio
$Z_{\rm s}/X_{\rm s}$ which was reduced from the value 0.0245
used in Model S, illustrated in \fref{fig:csqinv}, to 0.0165.

\begin{figure}
\begin{center}
\includegraphics[width=4.5in]{\fig/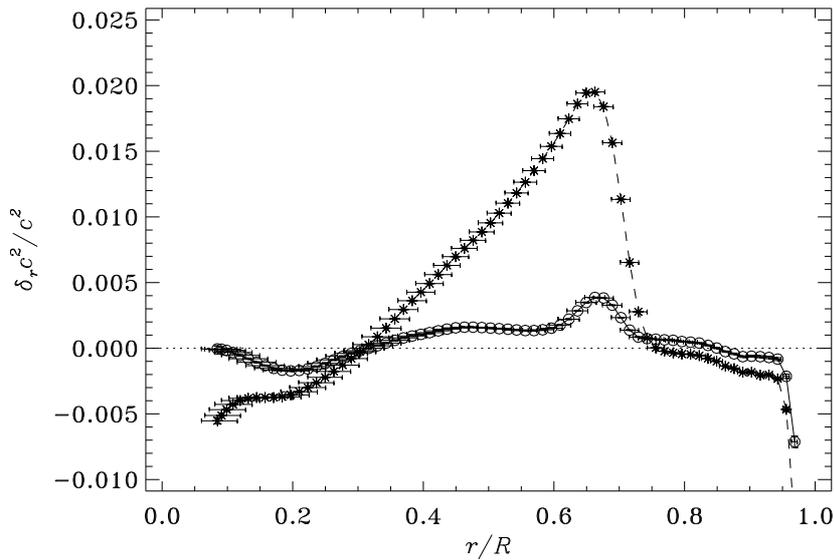}
\end{center}
\caption{Inferred differences in squared sound speed between 
the Sun and two solar models, in the sense (Sun) -- (model),
inferred from inversion of observed solar frequencies.
The open circles show results for Model S, as illustrated in
\fref{fig:csqinv},
while the stars are based on a model 
using the Asplund {\etal}\cite{asplundetal09} (AGSS09) composition.
See also the caption to \fref{fig:csqinv}.
}
\label{fig:agss09}
\end{figure}

This has dramatic consequences for solar modelling, discussed by 
Villante (these proceedings). 
Compared with Model S the maximum difference in $\delta_r c^2/c^2$ 
increased by about a factor of five.
A detailed review of the consequences for solar modelling and the
comparison with helioseismology was given by Basu \& Antia\cite{basu_antia08}.
Later developments of the techniques resulted in some increases in
the abundances, as reviewed by Asplund {\etal}\cite{asplundetal09} (AGSS09),
yielding $Z_{\rm s}/X_{\rm s} = 0.0181$,
still substantially below the original value.
The consequence for the model sound speed is illustrated in \fref{fig:agss09} 
which compares the inferred sound-speed difference for a solar model using the
AGSS09 composition with that for Model S.
Vinyoles {\etal}\cite{vinyolesetal17} carried out a detailed analysis of 
effects of the revised composition on the models.
An important point is that the effects on the predicted neutrino fluxes
is barely significant, as is also suggested by the modest differences
between the models and the Sun in the solar core, shown in \fref{fig:agss09}.
However, it is evident that the increased difference between 
the helioseismically inferred sound speed and the solar models represents
a serious problem for the modelling, which is also reflected in other seismic
properties of the Sun such as the depth of the convective envelope 
and the envelope helium abundance.
So far no definite solution to this problem has been found.

The heavy elements predominantly affect solar modelling through their 
contributions to the opacity,
and hence a possible solution to the discrepancy between the new models and
the Sun is to postulate errors 
in the opacity calculations.\cite{basu_antia04, montalbanetal04, bahcalletal05}
Indeed, it was shown in Refs \citenum{cdetal09, cd_houdek10} that with
a suitable opacity modification, assumed to depend just on temperature,
the structure of Model S could be recovered with the revised abundances.
In the case of AGSS09 this required a change of about 20 per cent at the
base of the convective envelope, decreasing smoothly to 2 per cent at
the solar centre.
However, these modifications are entirely {\it ad hoc}, and the important 
question is whether errors at this level in the opacity calculations 
are physically reasonable.
One possible measure of the uncertainties in the calculations are the
differences between independent calculations carried out under somewhat
different assumptions.
These differences are generally at most at a level of 5 per cent, with
somewhat larger differences in one case in the solar core, 
and hence do not match the modifications required to compensate for the
composition change. 
However, in an experiment under conditions approaching those
at the base of the solar convection zone Bailey {\etal}\cite{baileyetal15} 
found an iron absorption coefficient substantially higher than  
theoretically predicted, indicating inadequacies in the treatment of
atomic physics used in current opacity calculations.
These are unavoidably simplified, and the neglect of transitions or processes
broadening the energy levels would have a tendency to under-estimate the
opacity.
Even so, it would be a remarkable coincidence if the errors in the
opacity calculations were to match the effect of the increased abundances.

\begin{figure}
\begin{center}
\includegraphics[width=4.5in]{\fig/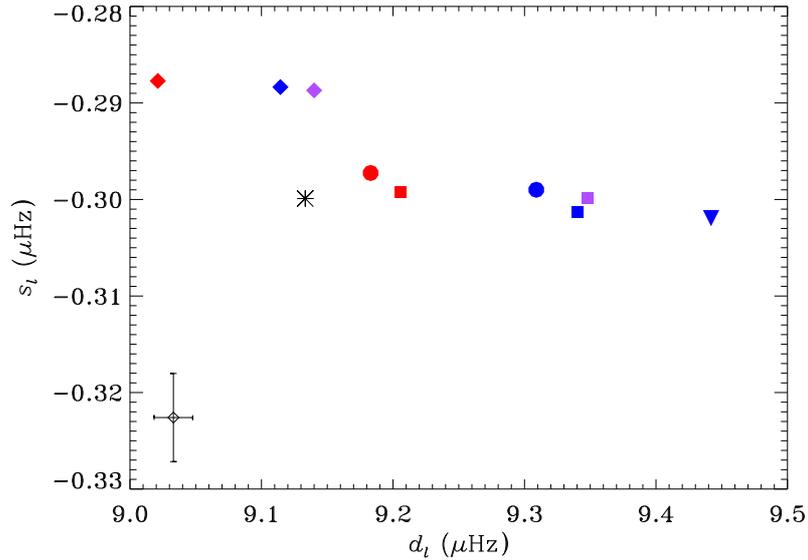}
\end{center}
\caption{Intercept $d_l$ and slope $s_l$ of the small separations 
$\delta \nu_{n0}$
fitted as a function of mode order ({\cf} Eq~\ref{eq:fitsep});
the fit included modes of radial order $n$ between 15 and 26 and used
a reference order $n_0 = 21$.
BiSON observations\cite{howeetal17} are shown by the diamond with error bars,
while the star shows results for Model S ({\cf} \fref{fig:csqinv}).
The remaining results were provided by Aldo Serenelli
(Villante et al., in preparation).
The symbol types are defined by the opacity tables:
circles are OPAL\cite{iglesias_rogers96}, 
squares are OP\cite{seaton_badnell04}, diamonds are OPLIB\cite{colganetal16}
and the triangle is OPAS\cite{mondetetal15}. 
The colour indicates the composition:
red for the Grevesse \& Sauval\cite{grevesse_sauval98} composition
and blue and purple for AGSS09\cite{asplundetal09} photospheric
and meteoritically corrected compositions (see Ref. \citenum{vinyolesetal17}).
Observations courtesy R. Howe; model results courtesy A. Serenelli.
}
\label{fig:fitsplit}
\end{figure}

To investigate the effects of the choice of composition and opacity table
on the core properties of the model, and to make a link to the analysis
of Elsworth {\etal}\cite{elsworthetal90}, \fref{fig:fitsplit}
shows the coefficients of the fit in \eref{eq:fitsep} to the
small frequency separation $\delta \nu_{n0}$, for recent observations and
for a selection of solar models.
Compared with the early analysis shown in \fref{fig:cdgafd91}
the observational errors have clearly been greatly reduced.
The results for the models cover a similar range as the standard
solar models in the original figure;
however, there are clear and significant differences between the
observed and computed values, which require further investigation.
In most cases the differences are smaller 
for the Grevesse \& Sauval composition than for AGSS09.
The exception are the OPLIB results where, on the other hand,
Serenelli (private communication) noted that the neutrino results
show substantial deviations from the observations.

\section{Concluding remarks}

At a personal level, solar neutrinos provided my way into what has been 
a scientific life dominated by helio- and asteroseismology. 
More broadly, both neutrino studies and helioseismology have 
seen remarkable developments over the past five decades, 
involving huge investments in increasingly sophisticated experimental
and observational equipment, with a parallel development of
increasingly detailed solar models.
As clearly demonstrated by the Dresden conference, the future
of neutrino physics is extremely promising, with a number
of very advanced facilities under development which will provide
detailed information about the solar neutrino spectrum.
In the case of helioseismology extensive observations are continuing
from the ground and from space, adding to the already very substantial
set of helioseismic data and, very importantly, following the subtle
variations in solar properties associated with the 11-year solar
magnetic cycle which may be undergoing substantial changes.\cite{howeetal17}
For helioseismic investigations
a major challenge is to optimize the techniques used to analyse the
helioseismic data, making full use of the data already available, 
to secure statistically reliable inferences of the structure and dynamics
of the solar core and other key parts of the Sun.

The present relation between solar neutrinos and helioseismology was
succinctly summarized by Haxton {\etal}:\cite{haxtonetal13}

\begin{quote}
{\it ``Effectively, the recent progress made on neutrino mixing angles
and mass differences has turned the neutrino  into a well-understood probe
of the Sun.
We now have two precise tools, helioseismology and neutrinos, 
that can be used to see into the solar interior.
We have come full circle: 
The Homestake experiment was to have been a measurement 
of the solar core temperature, until the solar neutrino problem intervened.''}
\end{quote}
With the challenges raised by the revisions of the inferred solar surface
composition we need to make full use of these two tools, in our continuing
attempt to understand the properties of the solar interior and the physics
that controls those properties.

\section*{Acknowledgments}

I am very grateful to Douglas Gough for introducing me to the study of
stellar oscillations and many other aspects of astrophysics, as well as
for friendship and fruitful collaboration over more than four decades.
I thank Yvonne Elsworth and Rachel Howe for providing up-to-date
data on low-degree solar oscillations, and Aldo Serenelli for results
on solar-model oscillation frequencies and for fruitful discussions of
aspects of solar neutrinos and solar oscillations.
The organizers are warmly thanked for the invitation to this conference
and the opportunity to renew contact with the field of solar neutrinos.
Funding for the Stellar Astrophysics Centre is provided by 
The Danish National Research Foundation (Grant DNRF106).


\end{document}